\documentclass[%
 aip,
 jmp,%
 amsmath,amssymb,
 reprint,%
]{revtex4-1}
\usepackage{graphicx}
\usepackage{dcolumn}
\usepackage{bm}

\begin{document}

v

\title{Binary mixture of hard disks as a model glass former: Caging and uncaging} 



\author{Sin-iti
Sirono}
\email[]{sirono@eps.nagoya-u.ac.jp}
\affiliation{Earth and Environmental Sciences, Nagoya University, Tikusa,
Nagoya 464-8601, Japan}

\date{\today}

\begin{abstract}
I have proposed a measure for the cage effect in glass forming systems. A binary
mixture of hard disks is numerically studied as a model glass
former. A network is constructed on the basis of the colliding pairs of
disks. A rigidity matrix is formed from the isostatic (rigid)
sub--network, corresponding to a cage. The determinant of the matrix
changes its sign when an uncaging event occurs. Time evolution of the
number of the uncaging events is determined numerically. I have found that
there is a gap in the uncaging timescales between the cages involving
different numbers of disks. Caging of one disk by two neighboring disks
sustains for a longer time as compared with other cages involving more than one
disk. This gap causes two--step relaxation of this system.
\end{abstract}

\maketitle 



%
%
\section{Introduction}

When a glass-forming material is cooled, the motion of the
constituents becomes extremely slow. Likewise, the mobility of a
granular material decreases as the grains are tightly packed. Although
the glass forming materials are thermal and the granular material is
athermal, many features similar to that of glass-forming materials
have been found in a granular system\cite{Reis}. This similarity
suggests a common mechanism in these systems. The motion of the
constituents is restricted due to their neighbors. This is called
'cage effect'\cite{Donth}. Clearly, the cage effect is of geometrical
origin, regardless of whether a material is thermal or athermal. If a constituent cannot escape from a cage after cooling or packing has taken place, the system freezes. This is observed as glass transition or jamming
transition for granular materials. The cage effect plays a critical
role in these transitions.

The cage effect has been studied using various procedures. A plateau
in the time evolution of the mean square displacement has been
found\cite{Donth,Binder} between the initial ballistic regime and the
later diffusive regime. A constituent rattles inside a cage during the
plateau. The intermediate scattering function displays a two--step
relaxation in accordance with the mean square
displacement\cite{Donth,Binder}. Cage correlation
function\cite{Rabani} defined by the list of neighbor's successfully
reproduces the temperature dependence of a diffusion constant.  Three point velocity correlation function captures the correlation of
directions of motion between different times, and shows a backward
motion of constituents caused by a cage\cite{Doliwa}. Non-Gaussian parameter reveals that the velocity distribution of the
constituents deviates upwardly at a high velocity range from a simple
Gaussian distribution, corresponding to a correlated motion induced by
a cage\cite{Weeks}. The four point correlation function\cite{Flenner}
shows the correlations between the relaxation of different
constituents. The jump of a constituent\cite{Vollmayr} can be divided into
two types: reversible and irreversible. The ratio of irreversible
to reversible jumps increases with temperature, corresponding to an
increase in the escape from a cage.

All of these functions or measures deal with the motion of the caged
constituents. The effects of a cage are discussed indirectly through
the motion of the caged constituents. There is no measure directly dealing
with the caging constituents adjoining to the caged
constituents. I have proposed a measure for the cage effect based on the
determinant of the rigidity matrix\cite{Graver}. The
concept of rigidity has been applied to network
glasses\cite{Phillips}, and fluid-solid transition\cite{Huerta}.

I have selected a mixture of hard disks, which has already been
investigated by many authors\cite{Alder,Speedy,Isobe}, as a model
glass former. The disks collide with each other elastically, without
energy dissipation.  As the packing fraction $\phi$ (fractional area
occupied by disks) increases , a two--step relaxation of the density
fluctuation\cite{Doliwa} is observed in this system. 

In the next section, I define a local network of interacting disks,
characterizing a cage. I introduce a parameter which quantifies the
caging in Section 3. Numerical methods are explained in Section
4. Numerical results are presented in Section 5. Section 6 gives
discussions and conclusions.

\begin{figure}
\includegraphics[width=7cm,height=5.2cm]{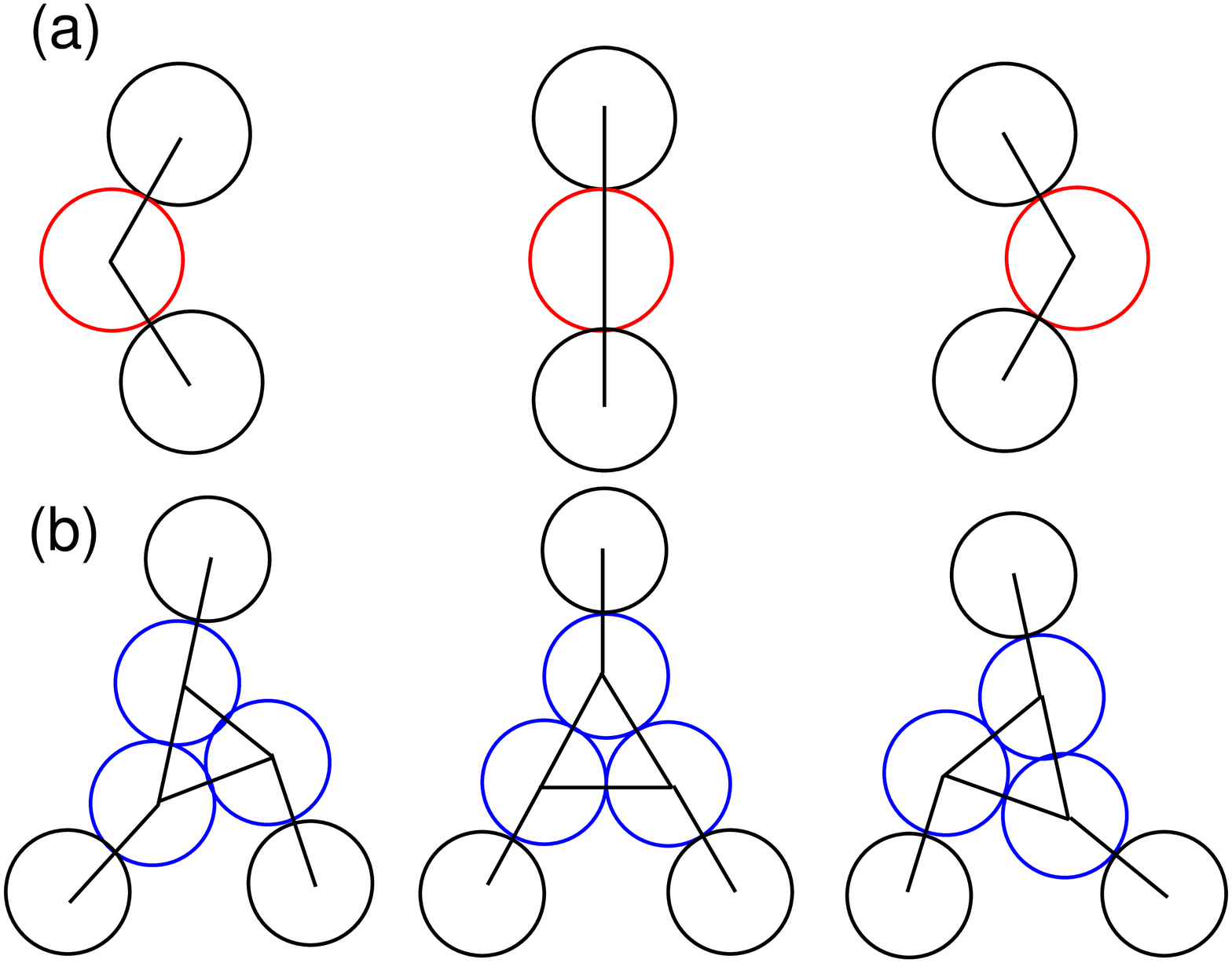}%
\caption{\label{fig1}(color online). Schematics of the cage effect. (a) Red disk is
  caged between two black disks. The red disk cannot pass through between
  the two disks (from the left to the right) unless the gap
  between the black disks is larger than the diameter of the red disk (middle).
Fictitious networks representing contacts between disks are shown by black lines.  (b) A cluster of three blue disks is caged by three black disks.
  }%
\end{figure}

\section{Locally rigid framework}

Figure~1 shows two examples of the cage effect. The red disk caged
between two black disks [Fig.~1(a)] cannot pass through between the two disks to
the other side unless the gap between the two disks increases up to
the diameter of the red disk. Likewise, a triangle of three disks
constrained by the three surrounding disks [Fig.~1(b)] cannot rotate
unless the top disk shifts upwards such that the two triangles, one of
which is formed by the centers of the three caged  disks, and the other
consists of the centers of the surrounding disks, are perspective
from a line (the extensions of three pairs of corresponding sides of
the two triangles meet in collinear points\cite{White}). 

If we apply an external force to the red disk to make it pass through
between the two caging disks or to the blue disks to rotate the
triangle, force carrying networks are formed as shown by the black
bonds in Fig.~1 provided that the positions of the end disks are
fixed. If we place hinges at the centers of the disks connected by
rigid bonds rotating freely around the hinges, the networks in Fig.~1
are rigid and isostatic\cite{Moukarzel}, such that any removal of a
bond results in a network that is not rigid. If the number of the
caged disk is $N_{\rm disk}$, the number of columns of the matrix is
$2N_{\rm disk}$ (there are two bonds in Fig.~1(a) ($N_{\rm disk}=1$)
and six bonds in Fig.~1(b) ($N_{\rm disk}=3$)). This is a property of
an isostatic network.

In the disk system, the combination of the colliding pair of the caged
disks does not change for a short timescale. From the list of
colliding pairs, we can construct a network where each bond represents
the colliding disk pair. If a disk (or a set of disks) is caged, we
can find a set of bonds corresponding to those shown in Fig.~1. This
is because the condition of uncaging shown in Fig.~1 is the same even
if the disks move and collide with each other. Therefore, we can define an
isostatic sub--network in the network as a cage. This local sub--network of
bonds  is called the 'locally rigid
framework (LRF)'. An LRF inhibits the motion of a disk (or a set of
disks) in one particular direction. If all LRFs pertaining to a
disk are effective, the disk can only rattle inside the
cage. It should be noted that this network is not a real contact
network, but a fictitious one because the disks do not continue to
contact but they collide. Moreover, an LRF may be broken as the system
evolves and distances between the disks increase.

\section{Caging parameter}

An uncaging event can be detected by the rigidity matrix\cite{Graver}.
We can construct the rigidity matrix from the LRF such that each row
represents a bond and each column corresponds to the coordinates of
caged disks. An LRF consists of bonds inside caged disks in addition
to bonds connecting between a caged disk and a neighboring caging
disk. There are two bonds connecting a caged disk (red) and caging
disks (black) in Fig.~1(a). For Fig.~1(b), there are three bonds
inside the caged disks (blue) and three bonds connecting the caged
disks and caging disks (black).

If there are bonds between the caged disks 1 -- 2 and 2 -- 3,
the rigidity matrix is written as\cite{Graver}
\begin{widetext}
\begin{eqnarray}
\begin{array}{c}
\\
\\
{\rm bonds}
\end{array}
\begin{array}{c}
\\
\\
1-2\\
2-3\\
\ldots\\
\end{array}
\begin{array}{c}
 {\rm coordinates\ of\ the\ caged\ disks}\\ 
\begin{array}{cccccccc}
\hspace{0.1cm}x_1\hspace{0.5cm} &\hspace{0.6cm}y_1\hspace{0.5cm} &\hspace{0.5cm}x_2\hspace{0.6cm} &\hspace{0.5cm}y_2\hspace{0.5cm} &\hspace{0.5cm}x_3\hspace{0.5cm} &\hspace{0.5cm}y_3\hspace{0.7cm}&\hspace{0.3cm} \ldots &  \\
\end{array}\\
\left(
\begin{array}{cccccccc}
{x_1-x_2\over d_{12}} & {y_1-y_2\over d_{12}} & {x_2-x_1\over d_{21}} &
 {y_2-y_1\over d_{21}} & 0 & 0 &\ldots &  \\
0 & 0 & {x_2-x_3\over d_{23}} & {y_2-y_3\over d_{23}} & {x_3-x_2\over d_{32}} & {y_3-y_2\over d_{32}} & \ldots &  \\
\multicolumn{7}{c}{\dotfill}\\
\end{array}
\right),
\end{array}
\label{rmat}
\end{eqnarray} 
\end{widetext}
where $x_i$ and $y_i$ are the positions of the $i$-th disk, and
$d_{ij}=\sqrt{(x_i-x_j)^2+(y_i-y_j)^2}$ is the distance between the
centers of disks $i$ and $j$. It should be noted that the matrix is
not symmetrical with exchanging $i$ and $j$. If $i$-th disk is caged
by $j$-th caging disk, the matrix contains $(x_i-x_j)/d_{ij}$ but does not
contain $(x_j-x_i)/d_{ij}$. 

The rigidity matrix for an isostatic network is a
square matrix (the number of degrees of freedom of the caged disks
$2N_{\rm disk}$ and the number of bonds are the same in an isostatic
network\cite{Graver}) and has a determinant except for a special
position of the disks\cite{Graver} when uncaging occurs.

When the red disk passes between the other two disks in Fig.~1(a), the
determinant of the LRF becomes zero because the two bonds are in
parallel when the three centers of the disks are collinear. The
determinant changes its sign at this point. Likewise, the sign of the
determinant in the right panel is different from that in the left
panel in Fig.~1(b).

Using the determinant of the LRF with $N_{\rm disk}$ caged disks,
we can define the caging parameter $\chi_{N_{\rm disk}}$, which is a measure of
the effectiveness of cages. It is defined as
\begin{equation}
\chi_{N_{\rm disk}}(t)=1-{\Pi_{N_{\rm disk}}(t)\over \Sigma_{N_{\rm disk}}},
\end{equation}
where, $\Sigma_{N_{\rm disk}}$ is the number of LRFs investigated which have $N_{\rm disk}$ caged disks,
and $\Pi_{N_{\rm disk}}(t)$ is the number of LRFs whose determinants change
their sign among $\Sigma_{N_{\rm disk}}$. It should be noted that the
number $\Pi_{N_{\rm disk}}(t)$ is increased by one when a determinant
changes its sign, only if an LRF has not been broken such that the lengths of
all the bonds $d_{ij}$ are less than $f(R_i+R_j)$, where $R_i$ is the radius of
the $i$th disk. The factor $f=1.5$ is adopted here.

\section{Numerical methods}

To determine the evolution of $\chi_{N_{\rm disk}}(t)$, the positions
of disks should be known. The evolution of the positions of hard disks
can be efficiently computed by an event-driven
simulation\cite{Isobe}, in which the collision times between disks and
velocities after collisions are sequentially solved. Disks having two
different radii, 1.0 and 1.4 in dimensionless units, are mixed in
equal proportions. This size ratio has been chosen to safely avoid
crystallization\cite{Speedy}. The total number of disks is 1024.  A
periodic boundary condition is applied. The packing fraction of the
disks $\phi$ is 0.8. The initial velocity distribution is Gaussian
with unit variance in dimensionless units.  I have defined the
colliding pair of the disks in a particular time window of 200 in the
normalized unit. The fictitious network is constructed from the list
of colliding disks in the time window.

\begin{figure}
\includegraphics[width=5cm,height=5.2cm]{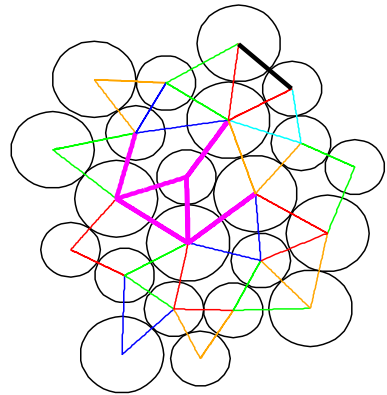}%
\caption{\label{fig2}(color online). An example of LRFs. Disks and bonds are chosen in a circular region in the computational area. Sets of bonds denoted by different colors are identified as LRFs on the basis of the rigidity of the local sets of bonds. An LRF of 6 bonds is denoted by the thick pink lines. The thick black line is the residual bond after decomposition}%
\end{figure}

In order to construct an LRF, a circular region is first placed
randomly on the fictitious network. The disks and the connecting bonds
inside the circle are chosen. By varying the radius of the circle, the
number of disks inside the circle is changed between three and hundred, which
is much smaller than the system size (1024 disks). The number of
chosen bonds is usually more than that required for an isostatic
network. This is because the disks move and have a chance to collide
with more disks than when they are tightly packed.

Overconstraining bonds (their removal does not affect the rigidity of
the network) are randomly removed until the isostatic
condition\cite{Jacobs} is fulfilled. An example of an isostatic
network is shown in Fig.~2. If a bond is further removed from the
isostatic network, some fraction of the bonds becomes
mobile. Correspondingly, a disk (or a set of disks) also becomes
mobile (rotates around a disk center). For example, if we remove a
blue bond at the bottom of Fig.~2, another blue bond becomes mobile
and the disk connected by this blue bond can rotate around. We can
find a minimum set of mobile bonds among $2N_{\rm disk}$ possible bond
removals. The mobile bonds can be found by the pebble game
algorithm\cite{Jacobs}.

In addition to the removed bond, the induced mobile bonds are further
removed from the isostatic network. In the example above, we remove
two blue bonds from the network. An LRF is composed of this local
mobile network plus the removed bond. The remaining network, after the
removal, is still isostatic. During each removal of bonds, the mobile
disk (caged disk) is also removed accordingly. The disks adjacent to
the mobile disk connected through the removed bonds correspond to the
caging disks. This removal process is repeated. Finally, the isostatic
network is uniquely decomposed into the sets of LRFs plus one bond
(thick black bond in Fig.~2).  In Fig.~2, there are 17 LRFs with 2
bonds ($N_{\rm disk}=1$) and 1 LRF with 6 bonds ($N_{\rm disk}=3$)
shown by pink lines. The pattern of LRFs shown in Fig.~2 changes if
the position of the randomly chosen circular region is shifted or the
random sequence of the removal of the overconstraining bonds is
changed.

According to this procedure, the LRF of $N_{\rm disk}=2$ is impossible
because we can always find an immobile bond for a particular removal of a
bond. An LRF of $N_{\rm disk}=2$ splits into two LRFs of $N_{\rm
  disk}=1$. 

\begin{figure}
\includegraphics[width=8cm,height=6cm]{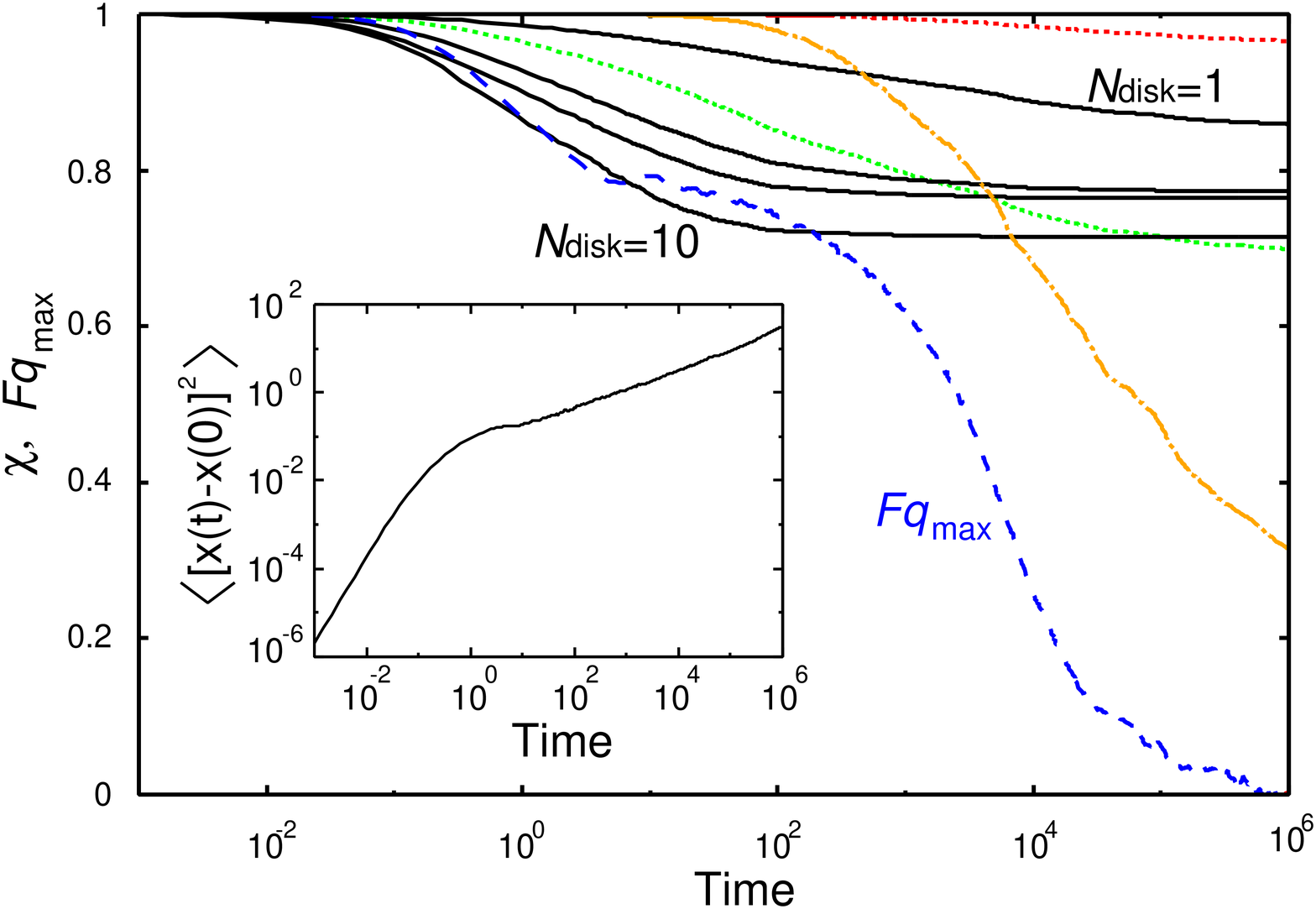}%
\caption{\label{fig3}(color online). Decrease in caging parameter
  $\chi_{N_{\rm disk}}(t)$ for various $N_{\rm disk}$ at
  $\phi=0.8$. From right to left, the black (solid) curves correspond
  to $N_{\rm disk}=1$, 3, 5, and 10. The red (upper dotted) curve
  represents the subset of $N_{\rm disk}=1$ in which a bond exists
  between the two caging disks when the bonds are placed according to
  the colliding pairs. The green (lower dotted) curve corresponds to
  the other group in which no bond is placed between the caging
  disks. The orange (dotted-dashed) curve is the same as the red (upper
  dotted) curve but $\Pi_{N_{\rm disk}}(t)$ in Eq.~(2) is multiplied by a factor of
  20. The blue (dashed) curve is the intermediate scattering function (self-part)
  $F_{q_{\rm max}}(t)$ at the wave number $q_{\rm max}$ where structure factor is maximum. [inset: mean square displacement]}%
\end{figure}

\section{Numerical results}

\subsection{Two--step relaxation}

Black curves in Fig.~3 display the evolution of $\chi_{N_{\rm
    disk}}$. From the right to the left of the black curves, the
number of caged disk $N_{\rm disk}$ is 1, 3, 5, and 10.  As $N_{\rm
  disk}$ increases, $\chi_{N_{\rm disk}}$ decreases faster. All
$\chi$s start to decrease at $t=10^{-2}$, when the intermediate
scattering function $F_{q_{\rm max}}(t)$ starts the first
relaxation. $F_{q_{\rm max}}(t)$ is given by
$\langle\Sigma_i\exp(\mbox{\boldmath$q$}_{\rm
  max}(\mbox{\boldmath$x$}_i(t)-\mbox{\boldmath$x$}_i(0)))\rangle$,
where $\mbox{\boldmath$x$}_i$ is the position of the $i$-th disk, the
angular bracket is the ensemble average, and $q_{\rm max}$ is the wave
number at which the structure factor is the maximum. This function
measures the relaxation of the structure with the nearest-neighboring
length scale. This first period corresponds to the ballistic motion of
the disks before a collision as can be seen in the inset of Fig.~3
where the mean square displacement, defined by $\langle
[\mbox{\boldmath$x$}_i(t)-\mbox{\boldmath$x$}_i(0)]^2\rangle$,
increases as $t^2$. The LRFs uncaged during this stage are essentially
ineffective, or very weak.

$F_{q_{\rm max}}(t)$ reaches a plateau when $t\simeq 4$. All $\chi$s
continue to decrease after $F_{q_{\rm max}}(t)$ reaches the
plateau. We cannot find any coincidence between $\chi$s (solid curves)
and the second relaxation of $F_{q_{\rm max}}(t)$.

The decrease in $\chi_3$, $\chi_5$, and $\chi_{10}$ ceases
when $t\simeq 10^2$. All LRFs belonging to these $\chi$s are uncaged
or broken by this time.

The decrease in $\chi_1$ (the rightmost black curve) is clearly
distinct from the others. There is a gap in the uncaging times between
$\chi_1$ and $\chi_3$. The uncaging time (the time when $\chi_{N_{\rm
disk}}(t)=0.9$) for $\chi_1$ is larger than that for $\chi_3$ by a
factor of $1.8\times 10^3$ (Fig.4(a)).  It should be noted that there
is no particular behavior in $\chi_1$ at $t\simeq 10^2$, when
$F_{q_{\rm max}}(t)$ starts its second relaxation.

This gap is emphasized if we divide the LRFs of $N_{\rm disk}=1$ into
two groups. In one group $\chi_{1,{\rm LO}}$, there is a bond between
the two caging disks, when the bond is placed on the basis of the
colliding pairs.  In the other group $\chi_{1,{\rm WK}}$, no bond is
placed between the caging disks (schematics are shown in
Fig.~4(b)). The decrease in $\chi_{1,{\rm LO}}$ and $\chi_{1,{\rm
    WK}}$ are shown as the red (upper dotted) and green (lower dotted)
curves respectively, in Fig.~3. The green curve starts to decrease at
the same period to other $\chi$s and continues to decrease through
$t\simeq 10^2$ without any notable change.

On the other hand, $\chi_{1,{\rm LO}}$ starts to decrease at $t\simeq
10^2$. Interestingly, this period coincides with the onset of the
second relaxation of $F_{q_{\rm max}}(t)$. There is no special signature at
$t=10^2$ in other $\chi$s. The amount of the decrease in $\chi_{1,{\rm
    LO}}$ is small, because more than 90\% of the LRFs investigated are
broken before uncaging. The orange (dotted-dashed) curve is the same as
the red curve, but $\Pi_{N_{\rm disk}}(t)$ in Eq.~(2) is multiplied
by a factor of 20 to make a comparison with $F_{q_{\rm max}}(t)$. A good
coincidence can be seen between the second relaxation of $F_{q_{\rm
    max}}(t)$ and $\chi_{1,{\rm LO}}$.

\begin{figure}
\includegraphics[width=8.4cm,height=12cm]{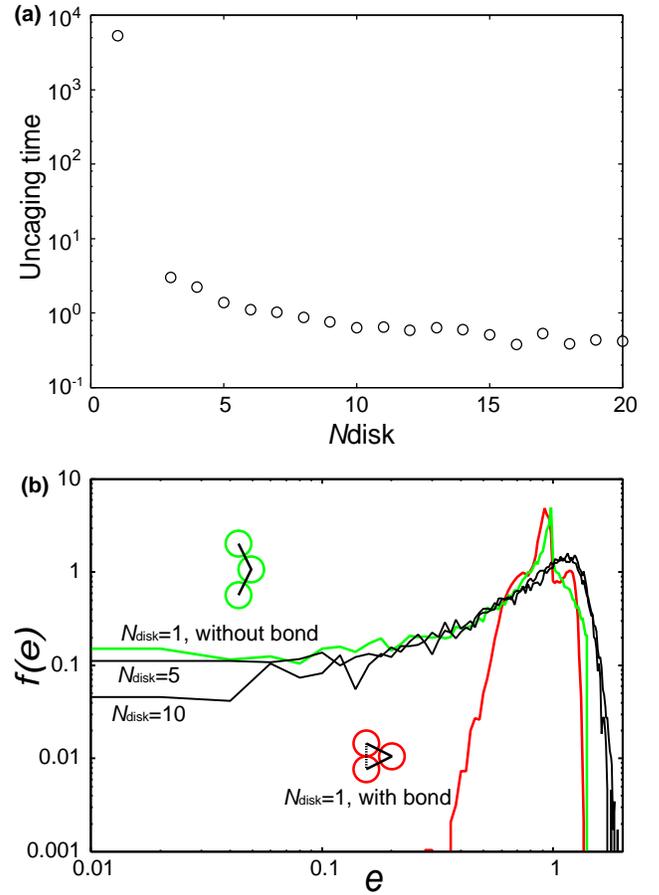}%
\caption{\label{fig4}(color online). Gaps between cages. (a) Uncaging
 time as a function of $N_{\rm disk}$. (b) The distribution of the
  absolute value of the eigenvalue $e$ of the rigidity matrix. The red
  curve stands for $N_{\rm disk}=1$ with a bond between the two caging
  disks, and the green curve for $N_{\rm disk}=1$ without the
  bond. The two black curves are for $N_{\rm disk}=5$ and $N_{\rm
    disk}=10$.}%
\end{figure}

\subsection{Gap between cages}

Figure~4(a) shows the uncaging time (the time when $\chi_{N_{\rm
    disk}}(t)=0.9$) as a function of $N_{\rm disk}$. It can be seen
that the uncaging time of LRFs with $N_{\rm disk}=1$ is significantly
larger than those of other $\chi$s. The uncaging time of other $\chi$s
decreases with $N_{\rm disk}$. A determinant $M$ of an LRF can be
factorized such that $M=HS$, where $H$ is composed of the bonds of an
isostatic subgraph (a triangle) in an LRF, and $S$ depends on the
pattern of the rest of the LRF\cite{White}. A determinant of an LRF
can be decomposed into factors corresponding to triangles and to
connecting bonds between triangles. A factor for a triangle cannot
change its sign because the LRF is broken before the
change. Therefore, the change in sign of a determinant for $N_{\rm
  disk}\ge 3$ comes from the factors of connecting bonds between
triangles. Because the number of such factors contained in $M$
increases as $N_{\rm disk}$, the uncaging time decreases as $N_{\rm
  disk}$.

The distribution $f(e)$ of the absolute value $e$ of a complex
eigenvalue of the rigidity matrix when $t=0$ is shown in Fig.~4(b). $f(e)$ is
defined such that $f(e)de$ is the fraction of the eigenvalues between
$e$ and $e+de$ and that $\int fde=1$. The eigenvalues evolve with
time followed by the evolution of the disk position. When one of the
eigenvalues of a rigidity matrix goes to zero, the determinant changes
its sign. Thus, the eigenvalue $e$ measures the difficulty of the
uncaging.  As $e$ increases, the displacements of the disks required
for uncaging increases. An eigenvalue corresponds to an eigenfrequency
if there is a harmonic potential between the disks. It can be seen that
there is a substantial fraction of low frequency modes for $N_{\rm
  disk}\ge 3$ in addition to $N_{\rm disk}=1$ without a bond. On the
other hand, the distribution $f(e)$ for $N_{\rm disk}=1$ with a bond
falls to zero at $e=0.3$. This difference is the origin of the gap in
uncaging times. An LRF with small $e$ uncages faster. However, an LRF
consisting of tightly packed disks sustains for a long time.

\section{Discussion and conclusion}

These observations suggest that the uncaging of LRFs of $N_{\rm
  disk}=1$ with a bond ($\chi_{1,{\rm LO}}$) requires the uncaging of
other LRFs. Uncaging of LRFs with $N_{\rm disk}\ge 3$ and $N_{\rm
  disk}=1$ without a bond proceeds during the plateau of the
intermediate scattering function. The LRFs relaxed in the first stage
flip over again between the neighboring configurations as shown in the
left and right panels in Fig.~1. The system gradually changes its
configuration by the flipping of the weak LRFs relaxed in the first
stage. The plateau shows that the disk system requires time to find a
configuration to open a path between two disks packed initially with
an order.

Displacements of disks due to the uncaging of these LRFs are small,
and can be seen as a plateau of the mean square displacement when
$t\simeq 10$, as shown in the inset of Fig.~3. This period corresponds
to $\beta$ relaxation of glass forming materials\cite{Binder}. Disk
positions gradually change in this period and the distance between two
caging disks in LRFs with $N_{\rm disk}=1$ with a bond, enlarges. The
uncaging of LRFs with $N_{\rm disk}=1$ with a bond triggers a large
structural change and promotes the second relaxation of $F_{q_{\rm
    max}}(t)$. This period corresponds to $\alpha$ relaxation of glass
forming materials\cite{Binder}.  The requirement of uncaging of other
types of LRFs for LRFs with $N_{\rm disk}=1$ with a bond is the
situation of ``constrained dynamics''\cite{Palmer} which explains the
slow relaxation of glass-forming systems. According to this study,
uncaging of LRFs $N_{\rm disk}=1$ with a bond cannot take place by
itself. Through the uncaging of LRFs with $N_{\rm disk}\ge 3$ and
$N_{\rm disk}=1$ without a bond, the arrangement of the disks changes
so that uncaging of LRFs $N_{\rm disk}=1$ with a bond takes place.

An LRF with $N_{\rm disk}=1$ with a bond is the collection of
locally ordered packing of three disks because the three disks collide
with each other when the network is determined. This local order
cannot propagate throughout because of the poly-dispersity of the
disks. This situation is called the geometrical
frustration\cite{Sadoc}. Because of the lack of the long-range order,
the geometrical frustration inevitably leads to the formation of
weak LRFs, which are uncaged first.

In conclusion, I have proposed a measure for the cage effect of
a binary mixture of disks. I have found the gap between cages with different
numbers of caged disks. Cages with one disk relax slower than those
involving more than one disk. The former cages correspond to the
locally ordered packing of disks. The latter cages represent
disordered packing of disks. This coexistence of locally ordered and
disordered disks leads to the two step relaxation.

\begin{acknowledgments}
\end{acknowledgments}

\providecommand{\noopsort}[1]{}\providecommand{\singleletter}[1]{#1}%

\end{document}